\documentclass[final]{aipproc}
\layoutstyle{6x9}

\begin{document}

\title{Decoding the Origin of Dark Matter}


\author{Zuowei Liu}
{address={C.N.\ Yang Institute for Theoretical Physics,
Stony Brook University, \\Stony Brook, New York 11794-3840, USA}
}

\begin{abstract}
We discuss the interplay between LHC signatures and the mechanism by which dark matter is generated 
in the early universe in supersymmetric theories. The LHC signatures of two of the major mechanisms 
for such generation of dark matter which are known to be the Stau Coannihilation (Stau-Co) region 
and annihilation on the Hyperbolic Branch (HB) are exhibited in detail. By analyzing the various 
LHC signatures, including multi leptons, hadronic jets, b-tagging, and missing transverse momentum, 
one can discriminate between the Stau-Co region and the HB region for the mSUGRA model. 
Interestingly, there are some regions of the parameter space which are beyond the current 
and near future reach of the dark matter direct detection experiments 
but will be accessible at the LHC, and vise versa.
\vspace{-8cm}
\begin{flushright}{YITP-SB-09-31}\end{flushright}
\vspace{7cm}
\vspace{1pc}
\end{abstract}

\maketitle

\section{Introduction}

Supersymmetry, and more specifically supergravity grand unified models, 
provide a well motivated 
framework for the exploration of new physics. Supergravity grand 
unified models also lead to the lightest neutralino as the lightest SUSY 
particle (LSP) which is a candidate for dark matter with R parity preserved. 

There are three broad regions in the parameter space of  supergravity models that satisfy the 
relic density measured by the WMAP  experiment \cite{Komatsu:2008hk}. 
These include (i) the Hyperbolic Branch (HB) where multi TeV scalars 
can appear consistent with small fine tuning (this region is alternatively 
referred as the Focus Point region) 
\cite{Chan:1997bi,Feng:1999mn}, 
(ii) The coannihilation regions 
where the coannihilation cross section between the LSP and the next 
lightest supersymmetric particle (NLSP) plays a significant role in 
satisfying the relic density, (iii) the Higgs pole region where dark matter 
annihilation cross section occurs near a Higgs boson pole. 
The coannihilation regions contain stau coannihilation \cite{Ellis:1998kh,Arnowitt:2001yh}, 
stop coannihilation, gluino coannihilation 
\cite{Profumo:2004wk,Feldman:2009zc,Gogoladze:2009bn}, etc. 
We focus our discussion on the stau coannihilation region (Stau-Co) and the HB region as 
these two are the more generic and also are the more probable models as 
suggested by the recent landscape analysis for different hierarchical 
mass patterns in mSUGRA 
\cite{Feldman:2007zn,Feldman:2007fq,Feldman:2008hs}.

\section{LHC signatures }

\begin{figure}[htb]
\includegraphics[width=7 cm,height= 6 cm]{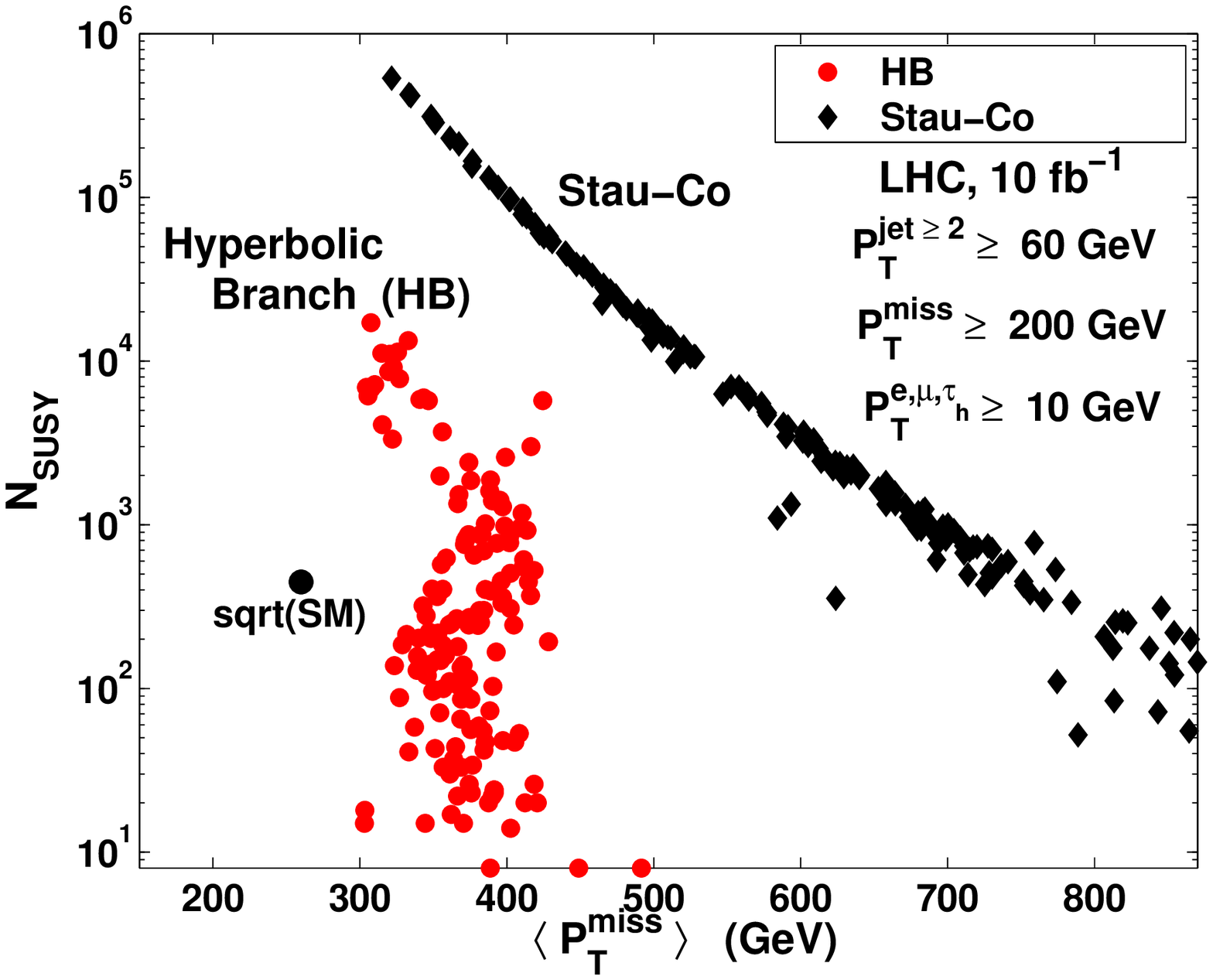}
\includegraphics[width=7 cm,height= 6 cm]{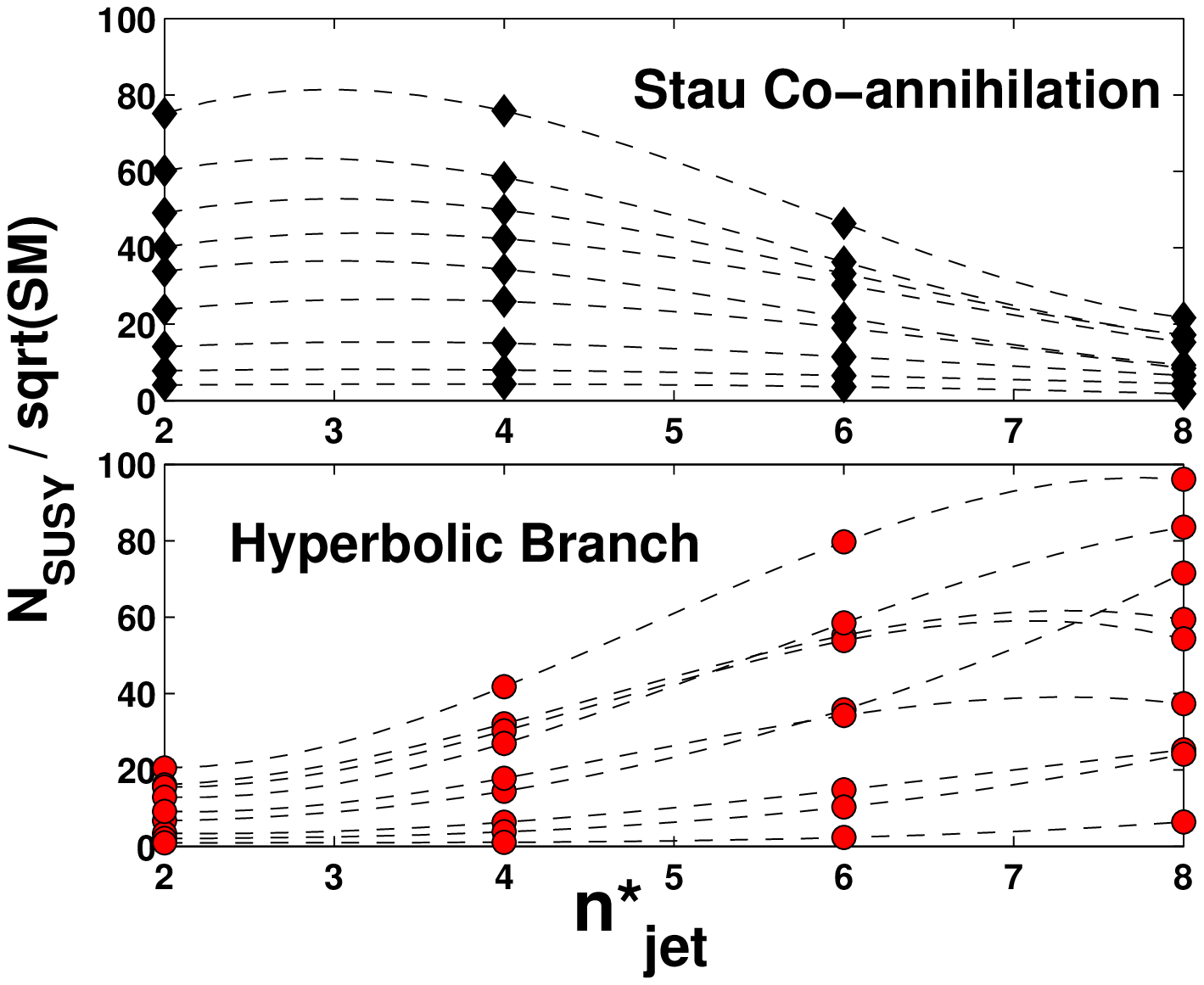}
\caption {
Left: $N_{\rm SUSY}$ vs.  $ \langle P^{\rm miss}_T \rangle$ 
for each model in the Stau-Co and HB regions.
 $ \langle P^{\rm miss}_T \rangle$ acts as an indicator of  Stau-Co and HB regions.
Right: Jet signature discrimination for HB and Stau-Co regions, where $n^{*}_{jet}$ is the 
minimal number of jets to pass the 
detector cuts. Figure taken from \cite{Feldman:2008jy}. }
 \label{fig:lhc}
\end{figure}

As mentioned above we focus our attention on 
the LHC signatures of the two major mechanisms for the 
generation of dark matter in the early universe, i.e., HB and Stau-Co. One of the most important 
LHC signatures of SUSY models is the missing transverse momentum, 
${\not\!\!{P_T}}$. A detailed analysis regarding the ${\not\!\!{P_T}}$ 
signature and of the total number of SUSY events 
is given in Fig.(\ref{fig:lhc}) where the average of the 
${\not\!\!{P_T}}$ is obtained for the events passing the post 
trigger detector cuts. 
Here one finds that $\langle{\not\!\!{P_T}}\rangle$ 
extends over a large range of energy scale for the Stau-Co models, 
while for the HB models this quantity lies within a much narrower band. 
Thus $\langle{\not\!\!{P_T}}\rangle$ can be viewed as a smoking gun 
signature to discriminate between the two mechanisms \cite{Feldman:2008jy}.

A complete quantitative analysis of the ${\not\!\!{P_T}}$ in an event 
is complicated due to the fact that it involves many particles 
and their decay chains   
\cite{Baer:1990sc,Bartl:1990ay,Djouadi:2000aq}. 
Here we give a qualitative discussion 
about this phenomenon. In the HB region, usually the gauginos are relatively 
lighter than squarks, and thus the SUSY productions at the LHC are dominated by
gaugino production. The copiously produced gluino then decays via an off-shell 
squark due to the hierarchical relations which results in a three body 
decay with two fermions and one gaugino. The other gauginos  
can also have dominant three body decay branching ratios which 
together make the cascade decay chains very lengthy. Thus the longer 
decay chains tend to produce a much reduced ${\not\!\!{P_T}}$ 
in the HB region.  
However, for the Stau-Co regions, the scalar mass is usually light 
and the squark production in the hadron collider becomes important. 
Because the squarks are usually lighter than the gluino in Stau-Co, 
the electroweak decay process with charginos or neutralinos as the final 
states can dominate the gluino final state. Further, $\tilde q_R$ can decay  
directly to the LSP and a quark, due to the gauge content of the neutralinos. Therefore, 
the decay chains of the Stau-Co models are much shorter which give 
rise to a relatively larger ${\not\!\!{P_T}}$.

We note that the HB models can also have many more jets in their LHC events 
than the Stau-Co models because of the difference in their cascade decays. 
Thus one can impose a large jet number cut for the purpose of 
optimizing the signal over the background in the HB region, 
while the same technique would overkill the signal in the Stau-Co region. 
A detailed analysis regarding this aspect is displayed in the Fig.(\ref{fig:lhc}).
This in turn also gives us a good discriminator between the two dark matter 
generation mechanisms.

Further, since in the HB region, the relatively lighter gluino has to undergo a three body 
decay via an off shell squark, the dominant modes are the ones with smaller 
virtuality which happen to be the lighter stop and the lighter sbottom. Thus 
the gluino decays are very rich in bottom quarks. This is again a remarkable 
signature for SUSY discovery using b-tagging and a great discriminator 
between the HB and the Stau-Co.

\section{Dark Matter Direct Detection}

\begin{figure}[htb]
\includegraphics[width=7cm,height=6cm]{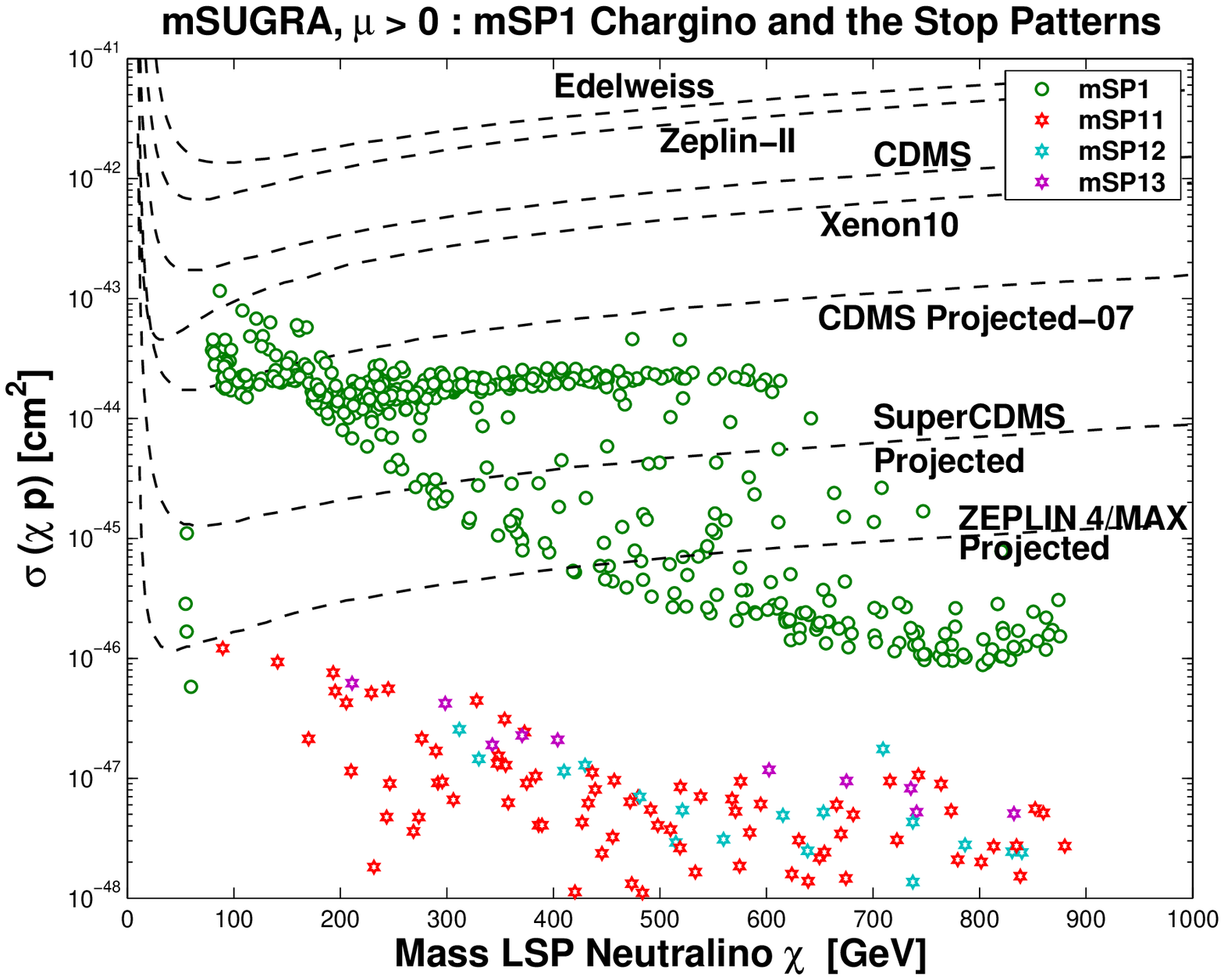}
\includegraphics[width=7cm,height=6cm]{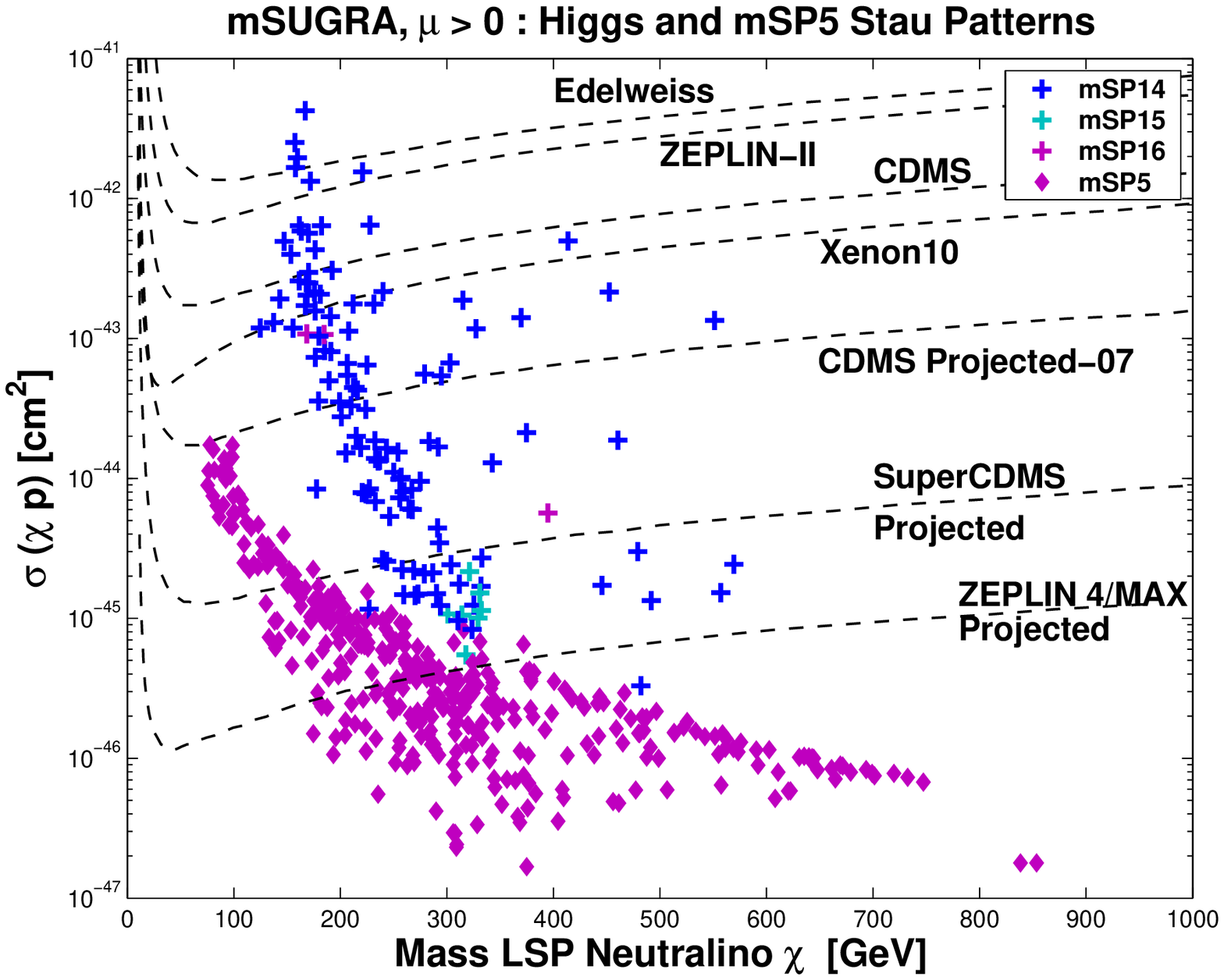}
\caption{Analysis of $\sigma(\chi p)$ in mSUGRA: 
(left panel)the Stau-Co models (mSP5) and Higgs patterns 
(mSP14-16); (right panel) the HB models (mSP1) 
and the Stop patterns (mSP11-13).  
A Wall consisting of models in the HB region 
with a $\sigma(\chi p)$ in the range $10^{-44\pm .5}$ cm$^2$
enhancing the prospects for the observation of dark matter by
SuperCDMS \cite{Schnee:2005pj},  ZEPLIN-MAX\cite{Atac:2005wv}
or LUX\cite{lux}. 
Figure taken from \cite{Feldman:2007fq}. }
\label{fig:dmdd}
\end{figure}

We discuss now the direct detection of dark matter. An analysis of the 
scalar neutralino-proton cross section $\sigma(\chi p)$ as a function 
of the LSP mass is given in Fig.(\ref{fig:dmdd}). We note that the models 
in the HB region and the Higgs pattern models 
give much larger dark matter cross sections than the ones 
in the Stau-Co region and the Stop patterns. The separation in 
this dark matter signature space makes it easy to identify the 
HB and Stau-Co mechanisms once the dark matter is discovered 
\cite{Feldman:2007fq}.

Another interesting phenomenon is the appearance of the Wall 
which runs horizontally to high LSP masses. 
The large neutralino-proton cross section 
arising from the Wall enhances the prospects for the discovery of 
dark matter for a broad range of energy scale. Although it is 
known that a large Higgisino component can give rise to strong 
neutralino-proton cross section, the finding that the Wall is composed of 
models arising mostly from the HB region is new and 
has not been observed before the work of \cite{Feldman:2007fq}.

\section{Dual Probes}

\begin{figure}[h]
\includegraphics[width=7cm,height=6cm]{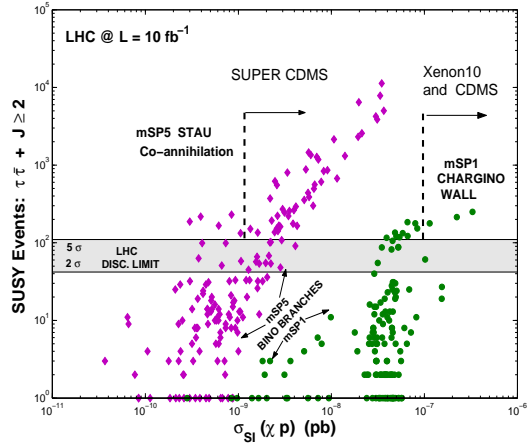}
\caption{An exhibition of the dual probes of SUSY by direct detection experiments and
by lepton, jet and missing energy signals at the LHC. The analysis above focuses on the
HB models (Chargino Pattern, mSP1) and 
the Stau-Co models (Stau Pattern, mSP5) for mSUGRA ($\mu > 0$). 
Figure taken from \cite{Feldman:2008en}. }
\label{fig:dual}
\end{figure}

Additional information regarding the discrimination between the HB and 
the Stau-Co regions can be obtained combining the LHC signature and 
the neutralino-proton cross section in the dark matter 
direct detection experiments. One finds that dark matter direct detection can 
complement the LHC search in some region of the parameter space. 
An example is given in Fig.(\ref{fig:dual}). Here one finds that a large 
collection of models originating from the HB region can be probed 
by the SuperCDMS experiment while they are unlikely to be 
discovered with 10 fb$^{-1}$ luminosity at the LHC. There are also 
models beyond the reach of the current and the near future 
dark matter experiments, but such models can be explored at the LHC. 
A clear separation in the plot of Fig.(\ref{fig:dual})
 again offers a way of distinguishing these two mechanisms, i.e., HB and Stau-Co.

\begin{theacknowledgments}
This work was supported in part by the NSF grants PHY-0757959 and PHY-0653342. 
\end{theacknowledgments}

\end{document}